\documentstyle[12pt,psfig]{article}
\begin{document}
\newcommand{\be}{\begin{equation}}
\newcommand{\bea}{\begin{eqnarray}}
\newcommand{\ee}{\end{equation}}
\newcommand{\eea}{\end{eqnarray}}
\newcommand{\lb}{\label}
\newcommand{\nl}{\newline}
\newcommand{\ra}{\rightarrow}
\newcommand{\nn}{\nonumber}
\newcommand{\al}{\alpha}
\newcommand{\de}{\delta}
\newcommand{\eps}{\epsilon}
\newcommand{\ga}{\gamma}
\newcommand{\la}{\lambda}
\newcommand{\om}{\omega}
\newcommand{\ns}{\normalsize}
\newcommand{\ph}{\phi}
\newcommand{\pr}{\prime}
\newcommand{\sld}{/\hspace{-0.55em}\partial}
\newcommand{\slk}{/\hspace{-0.55em}k}
\newcommand{\slp}{/\hspace{-0.58em}p}
\newcommand{\slq}{/\hspace{-0.5em}q}
\newcommand{\mi}{\mbox{i}}
\newcommand{\mt}{m_{top}}
\renewcommand{\thesection}{\arabic{section}.}
\renewcommand{\thefootnote}{\fnsymbol{footnote}}
\begin{titlepage}
\begin{quote}\raggedleft  {\bf LMU-14/93 } \\ {\em August 1993}\end{quote}
\vspace{1.5cm}
\begin{center}
{\bf LARGE N RELATIONS BETWEEN MASS MATRIX AND GAUGE GROUP
IN ELECTROWEAK GAUGE EXTENSIONS WITH STRONG COUPLING
\footnote{
to appear in the Proceedings of the 4th Hellenic School on Particle Physics}}
\end{center}
\vspace{1cm}
\begin{center}{\bf Ralf B\"{O}NISCH}

\vspace{1cm}
Ludwig-Maximilians-Universit\"{a}t M\"{u}nchen, Sektion Physik  \\
D -- 80333 M\"{u}nchen, Germany \\
\end{center}
\vspace{1cm}
\begin{abstract}
The Large $N$ expansion in gauge models with an NJL mechanism is discussed
in view of the resulting fermionic mass matrix. A hidden symmetry model is
introduced.
\end{abstract}
\vspace{2cm}
\end{titlepage}
\renewcommand{\thefootnote}{\fnsymbol{footnote}}
Electroweak models with an extendend gauge group
$SU(2)_L \times U(1)_Y \times G$ usually underly very strong conditions
like the need to match the
experimental $\rho$-parameter and other precisely known observables.
M. Lindner \cite{mali} reported on top condensate models, which use a
Nambu-Jona-Lasinio mechanism (NJL) to produce a condensate
$\langle \bar{t}t \rangle$ of the order the Fermi scale $G_F$ \cite{tc} .
To make some effort towards the physics beyond the SM,
they have recently been grounded on gauge extensions of the electroweak
sector, which necessarily have a strong coupling to trigger
the breakdown $SU(2)_L \times U(1)_Y \rightarrow U(1)_{em}$ \cite{models}.

Apart from the phenomenological and theoretical constraints which have
to be cared for in every extension of the gauge group, top condensate
models connect the fermionic mass matrix to the gauge group $G$.
Their interplay is strongly determined by the Large $N$ expansion, which
is used to treat the strongly coupled sector.
The parameter $N$ plays a crucial role
in the construction of the model and therefore carries information on the
possible structure beyond the SM.
We briefly recall the features of the Large $N$ expansion, as being
important here, and focus on the consequences for the condensation pattern.
Finally a specific model, which is motivated by substructure, will be
introduced.

In the classical NJL analysis of the interaction
$G  \bar{\psi}_L \psi_R \bar{\psi}_R \psi_L$,
fermion self-energies $\Pi(q^2)$
are summed to the effective scalar propagator as
\be \mi \Delta_{eff}(q^2)= \frac{\mi}{1- G \Pi(q^2)}
\lb{bubblesum},\ee
Fig. \ref{f1}.
The Schwinger-Dyson equation,
\bea
\delta m = m &=&  -G \Pi_{NJL} \nonumber \\
             &=& 2 G m \mi \int \frac{{\mbox d}^4p}{(2 \pi)^4}
                 \frac{1}{p^2 -m^2},
\lb{gap}\eea
is given by the graph in Fig. \ref{f1a}. It
is self-consistent by writing the renormalized mass $m$ also on the
righthand side. This includes graphs like Fig. \ref{f2}.

Restriction to a given order in $1/N$ allows analytical calculations
in limiting the number of conributing graphs \cite{witten}.
The validity of this approximation is the basic assumption of the NJL model.
Starting from an interaction
\be {\cal L}_{NJL}= \sum_{a,b=1}^N \left[ \bar{\psi}^a (\mi \sld -m) \psi_a
+\frac{G}{N} (\bar{\psi}^a_L \psi_{Ra})(\bar{\psi}^b_R \psi_{Lb}) \right],
\lb{sunnjl}\ee
graphs like Fig. \ref{f3} are of order 1/$N$ and suppressed in the limit
of large $N$:
The incoming and outgoing fermions carry the index $a$, it is the
mass counterterm $m_{aa}$.
Lines 1 and 2 in Fig. \ref{f3} have to be labeled $b$.
The summation over $b$ gives a factor $N_b$ for the graph.
Two of the lines 3, 4 and 5 carry an index $c$ and one carries an index
$b$, where the latter is fixed already
from the first vertex. Hence there is one factor $N_c$, but
no more factor $N_b$,
so that with a factor $1/N$ from each vertex the graph is of
order $N^2/N^3$.
Analogously, all
terms with an ingoing index $a$ and an outcoming
index $b$ or vice versa, i.e. mass counterterms $m_{ab}$ with $a \neq b$,
are suppressed in Large $N$.
Hence, starting with a strongly coupled
gauge group $G = SU(N)$ and expanding via this $N$ yields a mass matrix,
which is diagonal in the representation space of $G$.

In the summation eq.(\ref{bubblesum}), Fig. 1, side bubbles which are
connected to the chain by only one vertex are also taken into account by
using the renormalized fermion mass in the self-energy $\Pi(q^2)$,
while more complicated graphes are suppressed by orders of $N$.

The NJL graphs are the first order in $p^2/M^2$ of gauge theory ladder
diagrams, Fig. \ref{f4}, where
$p$ is a typical momentum and $M$ is the regulator of the NJL model or the
gauge boson mass.
The gauge vertex has to be index-changing to yield the dominant graphs of
the interaction eq. (\ref{sunnjl}):
The correct NJL-type interaction is
\be
{\cal L}_{eff} = - \frac{g^2}{2 M_A^2} \sum_{a,b=1}^N
\left(\bar{\psi}^a_L \ga_\mu \psi_{Lb} \right)
\left(\bar{\psi}^b_R \ga_\mu \psi_{Ra} \right),
\lb{njl}\ee
which
(due to the $t$-channel gauge boson exchange)
is Fierz-transformed into
\be {\cal L}_{eff} =   \frac{g^2}{M_A^2}
\sum_{a,b=1}^N
(\bar{\psi}^{a}_L \psi_{Ra})(\bar{\psi}^{b}_R \psi_{Lb}).
\lb{fierznjl}\ee
Hence, to obtain eq. (\ref{fierznjl})
from a gauge theory, one has
to start from the $SU(N)$ vertex
\be
{\cal L}_{gauge}= g \sum_{a,b=1}^N \bar{\psi}^a \ga_\mu \psi_b A^\mu_{ab}.
\lb{sungauge}\ee
In the Large $N$ expansion, the gauge coupling will be redefined
$g \rightarrow g/\sqrt{N}$.

In view of the various standard symmetries
a second fermionic $SU(N)$ degree
of freedom, which is independent from the first, will appear.
If the currents in eqn. (\ref{njl}) and (\ref{sungauge}) are corresponding
singlets,
\be
{\cal L}_{gauge}= g \sum_{a,b=1}^N \sum_{i=1}^M
\bar{\psi}^{a}_i \ga_\mu \psi_{b}^i A^\mu_{ab},
\lb{sungauge2}\ee
the effective scalar interaction takes the form
\be {\cal L}_{eff} =   \frac{g^2}{M_A^2}
\sum_{a,b=1}^N
\sum_{i,j=1}^M
(\bar{\psi}^{a}_{Li} \psi_{Ra}^j)(\bar{\psi}^{b}_{Rj} \psi_{Lb}^i).
\lb{fierznjl2}\ee
Now the Large $N$ expansion (i.e. summing independently over $a$ and $b$)
gives mass counterterms $m_{aa}^{ij}$ (and $m_{bb}^{ij}$), they are
constrained to be singlets only with respect to the Large $N$ index.
In the space of $i$ and $j$ the mass matrix has rank one,
there is only one nonvanishing eigenvalue \cite{rank1}.

What physical quantity should be represented by the parameter $N$?
The SM offers two $N$'s:
The color degree of freedom $N_c$ was used by C. Hill.
It might however be
unconvenient to give color such important meaning in the sector of
electroweak symmetry breaking.
The second candidate is the generation index $N_f$. Counting the number of
identical massless fermions,
it is tightly connected with the spectrum which we want to understand.
Consequently one can construct a strongly coupled horizontal gauge symmetry.
However, the mass matrix will be diagonal in flavor space and
a more complicated construction will be needed to produce a heavy top.
If the horizontal gauge bosons do not carry color,
$a, b$ in eq. (\ref{fierznjl2}) count generations and $i, j$ count color.
Thus, QCD might be broken simultaneously with $SU(2)_L \times U(1)_Y$.
This problem appears for all groups $G$, which do not embed $SU(3)_c$.
QCD corrections are supporting only color-singlet channels.
The strength of
their influence is mainly a question of the scale where the condensation
takes place.
It is not yet clear if and for which values of $M$ this problem can be cured.

One might speculate that fermion masses should not be the consequence
of external quantum numbers, but rather stem from a force which has to be
saturated within the massive objects. The mass hierarchies should then be
understood by inner dynamics.
This force would not directly show up at the Fermi scale and
yield a model with a hidden gauge symmetry, where
the parameter $N$ belongs to a hidden degree of freedom \cite{boekneur}.
We shall now describe such a model.

The breaking of local symmetries goes in two steps:
\be  \begin{array} {lcr}
SU(2)_L \times U(1)_{Y^\prime_L} \times U(1)_{Y^\prime_R} \times G &
\stackrel{spontaneously}{\longrightarrow}& \\
SU(2)_L \times U(1)_{Y_L} \times U(1)_{Y_R} &\stackrel{dynamically}
{\longrightarrow} & U(1)_{em}
\lb{break}\end{array}
\ee
The simplest Higgs-configuration $\Phi$ that works in the first step of eq.
(\ref{break}) transforms like $(0,\frac{1}{2},\frac{1}{2})$ with respect to
$SU(2)_L \times U(1)_{Y^\prime} \times G$.
For one neutral generator in $G$,
this $\Phi$ results in a mixing of the neutral sector:
\begin{equation}
\left( \begin{array}{c} W_{3} \\ v_{\mu}^{0} \\ B \end{array} \right) =
\left( \begin{array}{ccc} 1&0&0 \\ 0&\cos\xi& -\sin\xi \\ 0&\sin\xi&\cos\xi
                                                   \end{array} \right)
\left( \begin{array}{c} \bar{W}_{3} \\ \bar{V}_{3} \\ \bar{B} \end{array}
\right),
\lb{eq414}\end{equation}
where $\xi$ is given by the $U(1)_{Y^\prime}$ and $G$ gauge coupling constants
$g^\prime$ and $g^{\prime \prime}$,
\be \sin \xi = \frac{g^\prime}{\sqrt{g^{\prime \prime 2}+g^{\prime 2}}},
\label{mix} \ee
and $\bar{W_3}$, $\bar{B}$ and $\bar{V_3}$ are primordial neutral
gauge fields of
$SU(2)_L \times U(1)_{Y^\prime} \times G$.
$W_3$ and $B$ are massless standard gauge fields and $v^0$ is a new neutral
massive boson field.
Its mass is related to the VEV $v$ of $\Phi$ via
$m_v = \frac{1}{2} \sqrt{g^{\prime\prime 2}+g^{\pr 2}}$.
The only difference between $U(1)_{Y^\pr}$ and $U(1)_{Y}$
is in the gauge coupling constant (before and after mixing with $G$).
The interaction is given as
\bea {\cal L}_{I} &=& g J^{\mu}_{i} \bar{W}_{\mu i}
                    + g^\pr J^{\mu}_{Y} \bar{B}_{\mu}  \nn \\
                  &=& g J^{\mu}_{i} W_{\mu i}
                    + g^\pr \cos\xi J^{\mu}_{Y} B_{\mu}
                    - g_{v^0} J^{\mu}_{Y} v_{\mu}^{0},
\lb{ww}\eea
where $J_{i}$ ($J_{Y}$) is the standard isospin (hypercharge) current
and $g_{v^0}\equiv g^\pr \sin \xi$.
Charged bosons of $G$ do not couple, because there is no mixing with $SU(2)_L$.

Now we consider the model at low momentum transfers $q^2 \ll m_v^2$.
The effective contact interaction is
\bea
{\cal L}_{4F}=&-&\frac{g^\pr \sin\xi)^{2}}{2M_{v}^{2}}
J_{Yi}^{\mu} J_{Yj\mu}      \\
 =&+&\frac{G^{\alpha\beta}_{LL}}{M_{v}^{2}}
 \left( \bar{\psi}^{i\alpha}_{rL} \gamma^{\mu} \psi_{i\alpha L}^{r} \right)
 \left( \bar{\psi}^{j\beta}_{sL} \gamma_{\mu} \psi_{j \beta L}^{s} \right)
\nn\\
 &+&\frac{G^{\alpha\beta}_{RR}}{M_{v}^{2}}
 \left( \bar{\psi}^{i\alpha}_{rR} \gamma^{\mu} \psi_{i\alpha R}^{r} \right)
 \left( \bar{\psi}^{j\beta}_{sR} \gamma_{\mu} \psi_{j \beta R}^{s} \right)
\nn\\
 &-&\frac{G^{\alpha\beta}_{LR}}{2M_{v}^{2}}
 \left( \bar{\psi}^{i\alpha}_{rL} \gamma^{\mu} \psi_{i\alpha L}^{r} \right)
 \left( \bar{\psi}^{j\beta}_{sR} \gamma_{\mu} \psi_{j \beta R}^{s} \right)
+h.c.,
\label{ia}
\eea
where $i,j=1,2,3$ count generations, $r,s=1,2,3$ count colors and
$\alpha, \beta=u,d$ label isospin components.
The quantum numbers $Y_L$ and $Y_R \equiv Q$ are absorbed into the couplings
$G^{\al\beta}$, they are proportional to
\be Y_{L} \cdot Q= \left\{
\begin{array}{ll}\frac{1}{6}\cdot\frac{2}{3} &\mbox{for $I_{W_3}=1/2$ quarks}
\\
            \frac{1}{6}\cdot(-\frac{1}{3}) &\mbox{for $I_{W_3}=-1/2$ quarks.}
\end{array} \right . \lb{quarkhier}\ee
After Fierz-transformation eq. (\ref{ia}) reads
\bea
{\cal L}_{4F}=&+&\frac{G^{\alpha\beta}_{LL}}{M_{v}^{2}}
 \left( \bar{\psi}^{i\alpha}_{rL} \gamma^{\mu} \psi_{j \beta L}^{s} \right)
 \left( \bar{\psi}^{j\beta}_{sL} \gamma_{\mu} \psi_{i\alpha L}^{r}\right) \nn\\
 &+&\frac{G^{\alpha\beta}_{RR}}{M_{v}^{2}}
 \left( \bar{\psi}^{i\alpha}_{rR} \gamma^{\mu} \psi_{j \beta L}^{s}\right)
 \left( \bar{\psi}^{j\beta}_{sR} \gamma_{\mu} \psi_{i\alpha R}^{r}\right) \nn\\
 &+&\frac{G^{\alpha\beta}_{LR}}{M_{v}^{2}}
 \left( \bar{\psi}^{i\alpha}_{rL}  \psi_{j \beta R}^{s} \right)
 \left( \bar{\psi}^{j\beta}_{sR}   \psi_{i\alpha L}^{r} \right) +h.c.
\label{fierzia}
\eea
As discussed above, the mass matrix is of rank one in generation space and
according to eq. (\ref{quarkhier}) only $I=+1/2$. quarks condense, while
the interaction is repulsive for $I=-1/2$ quarks. Therefore, among quarks
only the top is massive. This mass splitting is a consequence of the
$SU(2)_R$ violation of $U(1)_{Y_R}$ quantum numbers.

We have not adressed the compelling question of quadratic divergencies
and experimental compatibility of the exact value of $m_{top}$ here
\cite{bhj}.

Within the hidden symmetry model, there are several further
questions to be answered. First, there is a problem with the leptons,
because in a three-family scheme the $\tau$ gets very heavy. Going to a fourth
family avoids this problem, but the original idea was to explain the heaviness
of the top, which is extraordinary because it is the only fermion that is not
light on the Fermi scale.

Eqn. (\ref{mix}) and (\ref{ww}) tell us that for a large $g_{v^0}$,
$g^\pr/g^{\pr\pr}$ must be large.
By the usual running of gauge coupling constants,
this ratio is increasing with increasing energy and the situation
of coming down from a symmetric phase at a high scale looks troublesome,
at least if dim $G >$  1.

The questions of leptons and $SU(3)_c$-conservation also ask for the number $N$
of
hidden degrees of freedom. (Note, that this $N$ has nothing to do with the
dimension of $G$.) Deriving conditions (like e.g. lower bounds) on $N$
potentially gives hints on substructure and/or further physics beyond the
SM.

\vspace{2cm}
{\bf Acknowledgement} \newline
I thank the organizers very much for the invitation to this
communacative and inspiring workshop.

\newpage
\begin{figure}
\hspace*{0.5cm}
\psfig{figure=f1.eps,height=2cm,xscale=1.1,yscale=1.1}
\psfig{figure=f2.eps,height=2cm,xscale=1.1,yscale=1.1}
\caption{NJL-summation of fermion loops.}
\lb{f1}
\end{figure}

\begin{figure}\hspace*{3.75cm}
\psfig{figure=f3.eps,height=2cm,xscale=1.3,yscale=1.3}
\caption{NJL-self-energy entering eq. (\protect\ref{gap}).}
\lb{f1a}\end{figure}

\begin{figure}\hspace*{3.7cm}
\psfig{figure=f4.eps,height=3cm,xscale=1.3,yscale=1.3}
\caption{By self-consistency of eq. (\protect\ref{gap}) included graph.}
\lb{f2}\end{figure}

\begin{figure}[htb]
\hspace*{5.2cm}
\psfig{figure=f5.eps,height=4cm,xscale=1,yscale=.75}

\hspace*{3.1cm}
\psfig{figure=f6.eps,height=2.5cm,xscale=1,yscale=1}
\caption{Simple graph of order $1/N$.}
\lb{f3}
\end{figure}

\begin{figure}[htb]\hspace*{1.9cm}
\psfig{figure=f7.eps,height=3cm,xscale=1,yscale=1}
\hspace*{1.9cm}
\psfig{figure=f8.eps,height=4.5cm,xscale=1,yscale=1}
\caption{Ladder diagram with notation of the index $N$.}
\lb{f4}\end{figure}

\end{document}